\documentclass[a4]{article}

\usepackage[pdftex]{graphicx}

\usepackage{url}
\usepackage{amsmath}
\usepackage{authblk}

\newcommand{\fref}[1]{Fig.~\ref{#1}}
\newcommand{\fsref}[1]{Figs.~\ref{#1}}
\newcommand{\Fref}[1]{Figure~\ref{#1}}
\newcommand{\tref}[1]{Tab.~\ref{#1}}
\newcommand{\Tref}[1]{Table~\ref{#1}}
\newcommand{\eref}[1]{Eq.~(\ref{#1})}
\newcommand{\esref}[1]{Eqs.~(\ref{#1})}

\newcommand{\Esref}[1]{Equations~(\ref{#1})}
\newcommand{\scref}[1]{Sec.~\ref{#1}}
\newcommand{\SCref}[1]{Section~\ref{#1}}

\usepackage{bm}

\title{Analytical voltage estimation in power packet dispatching network}

\author[1]{Shinji Katayama}
\author[2]{Takashi Hikihara}

\affil[1,2]{Department of Electrical Engineering, Kyoto University\\
Katsura, Nishikyo, Kyoto 615-8510, Japan}
\affil[1]{s-katayama@dove.kuee.kyoto-u.ac.jp}
\affil[2]{hikihara.takashi.2n@kyoto-u.ac.jp}
\affil[ ]{\thanks{This paper was submitted to Nonlinear Theory and Its Applications, IEICE on December 28, 2021}
\thanks{This work was supported in part by Cross-ministerial Strategic Innovation
Promotion Program from New Energy and Industrial Technology Development
Organization, by Program on Open Innovation Platform with Enterprises, Research Institute and Academia (OPERA), by WISE Program, MEXT, and by JSPS KAKENHI under Grant 19J20591.}}

\begin{document}


\maketitle
\begin{abstract}
This paper proposes an analytical voltage estimation method for the power packet dispatching network.
A unit power packet consists of signals and DC pulsed voltage waveform.
In the network, power packets are transmitted among power packet routers (routers) by time-division multiplexing.
Routers have capacitors as the storage of power during the power transmission.
The power flows depending on the voltage gradient among the storage.
The power flow in the network has been studied with numerical approaches.
However, the existing studies have considered transmission losses but ignored transient responses.
Considering transient responses, the power transmission analysis generally requires circuit simulation and takes considerable time.
This paper focuses on a one-to-one connection between routers and underdamped conditions.
They allow the estimation of storage voltages considering the transient responses without circuit simulation.
First, theoretical analysis and experimental verification are performed on the conditions.
Then, an analytical voltage estimation method is proposed by expanding the analysis of the one-to-one connection to the network.
Comparison with circuit simulation reveals that the estimation method is accurate enough to analyze power transmission in the network.
\end{abstract}

\section{Introduction}\label{sec:intro}

Power packet dispatching system is a promising candidate for electrical energy transmission and management systems.
In the 1990s, Prof. Toyoda's research group proposed the power packetization concept \cite{Toyoda1998}.
Their proposal tried to simplify the trading of electric energy.
They proposed a virtual integration of power and related information with the accompanying network.
A unit of a specific amount of energy and its related information were traded like an IP packet on the Internet in their concept.
The physical integration of power and information was proposed in the 2010s \cite{Takuno2010,Abe2011a}.
The information is tied with the pulsed power as a series of rectangle voltage waveforms by the switching of wide-gap semiconductors \cite{Takuno2010}.
The physical integration enables simultaneous power and information transmission.
The power packet is transmitted by time-division multiplexing (TDM).
TDM allows power packets with different voltages to be dispatched on the same power lines.
Sources and loads also can share power converters.
TDM power transmission requires dead time to protect sources from undesirable high voltages.
The dead time can be utilized for other purposes.
Communication and sensing for fault detection are examples \cite{Takuno2010,Eaves2012}.
The power packet dispatching network focuses on communication and applies to energy management.

The power packet dispatching network has been studied with predominantly numerical and experimental approaches.
In numerical approaches, the consensus dynamics were applied to examine power transmission and control \cite{Ando2016b,Baek2020,Baek2020a}.
They assumed that power lines were resistive.
When the transient response of power transmission is negligible, the assumption is acceptable.
Preceding experimental studies have developed power packet routers (routers) and verified the power packet transmission \cite{Takuno2010,Takahashi2015a,Mochiyama2019,Naomitsu,Katayama2020a}.
Routers have capacitors as storage and generate power packets with the stored electric energy.
The power flow depends on the voltage gradient between routers and impedance \cite{Mochiyama2019,Katayama2020a}.
In these experimental studies, the assumption above is not always satisfied.
The effects of transient response appear in the current waveforms and transmitted energy.
Therefore, there is a gap between the assumed system in numerical studies and experimental systems.

This paper examines the power transmission considering the transient response.
Analysis considering the transient response generally requires circuit simulation.
It can take a considerable time on the network with multiple circuit elements like sources, routers, and loads.
This paper analyzes the power transmission by assuming a one-to-one connection of routers and underdamped conditions.
The one-to-one connection allows theoretical power transmission analysis considering the transient response.
The underdamped condition reduces the switching losses not considered in the analysis.
Combining these conditions allows estimating storage voltages and reconstruction of voltage waveforms.
The estimation method enables power transmission analysis considering the transient response.
In \scref{sec:packet}, the power packet dispatching network is explained.
\SCref{sec:trans} gives the theoretical analysis of power transmission in the network.
Power transmission in the network is examined in \scref{sec:estim}.
The experimental verification of power transmission with the underdamped condition is also given.
\SCref{sec:conclusion} is the conclusion of this paper.

\section{Power packet dispatching network}\label{sec:packet}

\begin{figure}[t]
 \centering
 \includegraphics[width = .6\hsize]{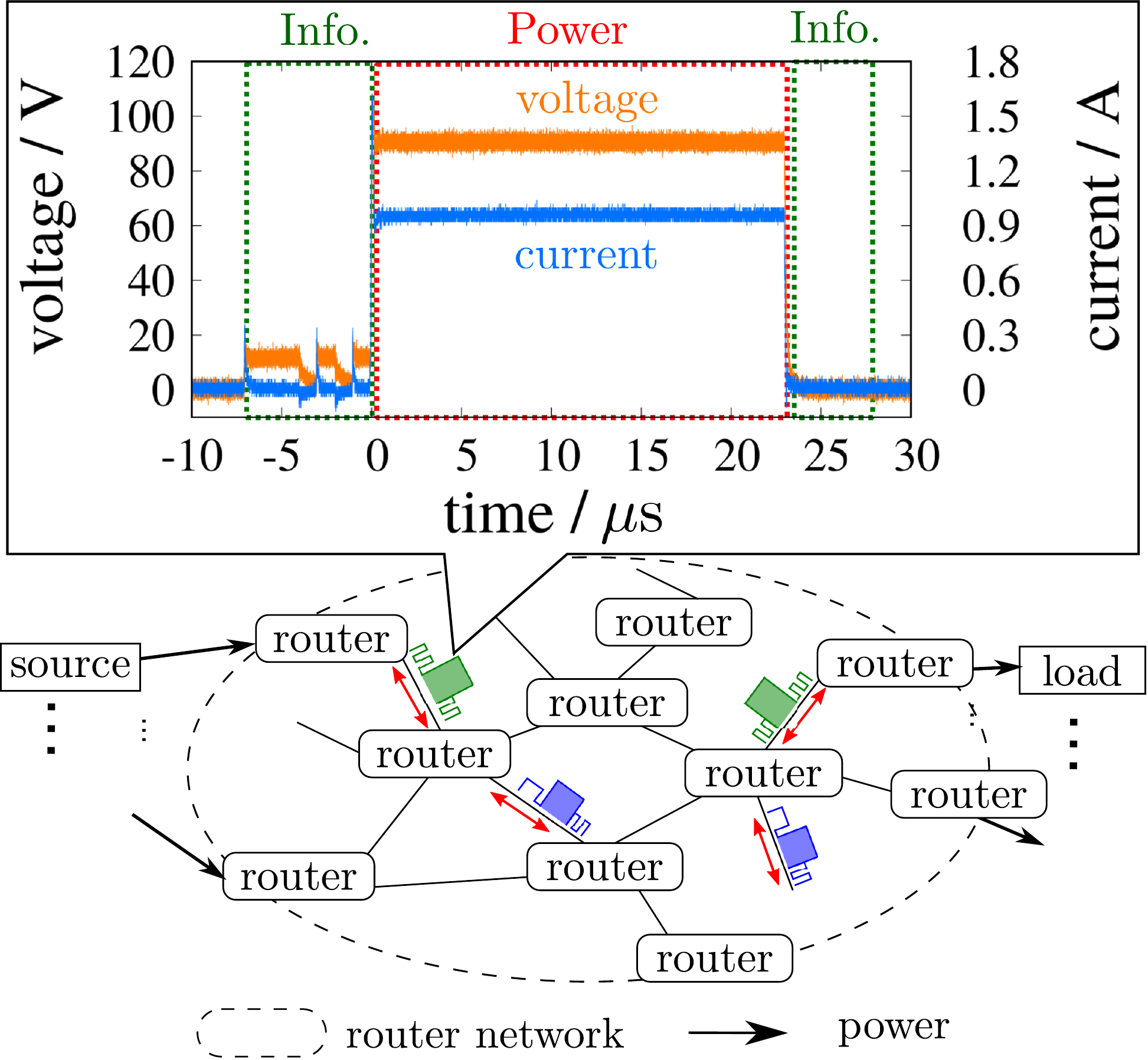}
 \caption{Concept of a unit power packet and power packet dispatching network.}
 \label{fig:concept}
\end{figure}

\begin{figure}[t]
 \centering
 \includegraphics[width = .7\hsize]{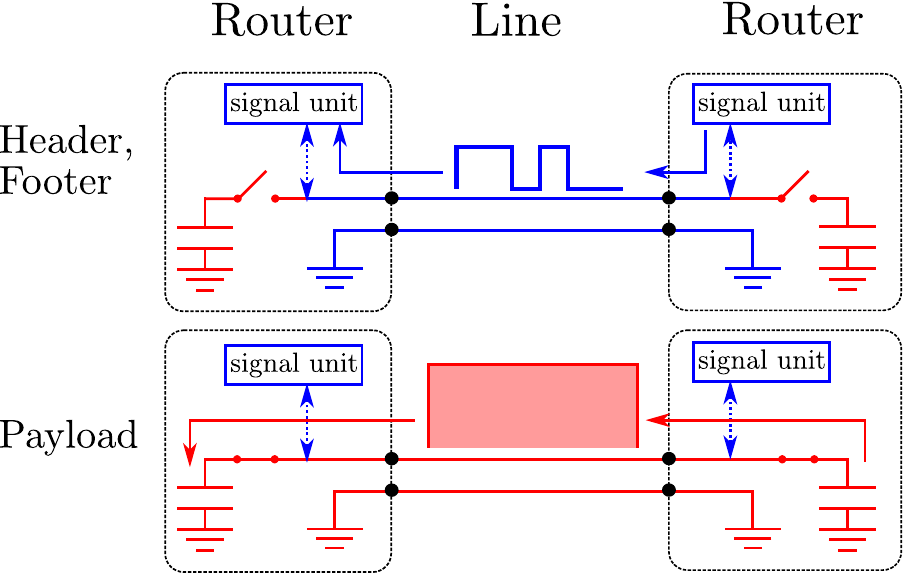}
 \caption{Schematic of power packet router and power packet transmission scheme.}
 \label{fig:router}
\end{figure}

A power packet is a unit of power and information.
Physically integrated power packets are discussed in this paper.
\fref{fig:concept} shows the power packet dispatching network \cite{Takuno2010}.
It shows the example waveforms of a unit power packet.
A unit power packet follows the configuration of IP packets on the Internet.
The header and footer send the information by a sequence of the rectangle voltage waveforms.
The payload transmits the electric power by the pulsed voltage waveforms with the corresponding current.
Power packet routers (routers) transmit power packets by TDM.
A router has a switching circuit, controller, and storage.
The switching circuit transmits the payload with the electric energy on the storage.
\fref{fig:router} depicts a router and power packet transmission scheme \cite{Katayama2020a}.
When a router (sender-router) starts transmitting a power packet to a neighboring router (receiver-router),
the sender-router generates the header signal.
The receiver-router reads the signal and recognizes the attributions of the following payload,
such as the voltage and time duration.
Then, the sender and receiver routers turn on the designated switches simultaneously and transmit power by a payload.
The footer signal follows the payload to let routers end a power packet transmission process.
The storage device is assumed to be a capacitor.
This paper refers to the prior experimental studies \cite{Mochiyama2019,Katayama2020a} and employs them.
The appropriate storage device depends on the amount of energy to hold.
The capacitor suits the payload with up to some kW and shorter than ms.

Power transmission in the power packet dispatching network has been analyzed with numerical approaches.
Power is a product of voltage and current.
Voltage is predominantly assumed to have rectangle waveforms \cite{Takuno2010,Ando2016b,Baek2020,Baek2020a,Gelenbe2016,Ma,Mochiyama2019}.
Some numerical studies have adopted a stable current waveform \cite{Gelenbe2012,Ma}.
It implies that the amount of transmitted energy linearly depends on the duration of power transmission.
The linear relationships between the amount of transmitted energy and transmission duration are seen in the power transmission to the resistive loads and converters.
The relationships allow the application of methods and theories in information networks to the power distribution system \cite{Gelenbe2016,Ma}.
However, the stable current is difficult to realize in the power transmission among capacitors.
Consensus dynamics are also applied for the analysis \cite{Ando2016b,Baek2020,Baek2020a}.
Their assumed systems have the capacitor as the storage device and the resistive power line.
A resistive power line assumption is acceptable when the transient response is negligible.
These studies revealed the characteristics of power distribution and control.

In experimental studies, rectangle voltage waveforms have been achieved by the hard switching of the wide-gap semiconductors \cite{Takuno2010,Takahashi2015a,Nakano2018,Mochiyama2019,Katayama2020a}.
The current depends on the voltage gradient and impedance \cite{Takuno2010,Takahashi2015a,Nakano2018,Mochiyama2019,Katayama2020a}.
The current is not always stable during the transmission, and transient response appears, especially in the short pulses.
The analysis of power transmission considering the transient response generally requires circuit simulations.
However, the circuit simulation takes considerable time in a network with many circuit elements including sources, routers, power lines, and loads.
To analyze an extensive network with multiple components, there is a need for an analysis method on power transmission without the circuit simulation.

Existing studies have revealed that the power transmission on a short power line does not experience severe performance deterioration \cite{Kajiyama2013,Takahashi2016a}.
Here a short power line means that the distributed elements model is unnecessary.
Based on the results, this paper assumes that the power lines are short enough to be modeled by a lumped constant model.

\section{Theoretical analysis of power transmission in power packet dispatching network}\label{sec:trans}

This section gives the theoretical analysis of power transmission in a power packet dispatching network.
The study begins with the power transmission on a one-to-one connection of routers.
Then, the analysis is expanded to the power transmission on multiple connections.
Finally, among the power transmission conditions, the one-to-one connection is found to have an analytical solution for power transmission.

\subsection{Analysis on power transmission between two routers}\label{sec:analysis}

\begin{figure}[t]
 \centering
 \includegraphics[width = .8\hsize]{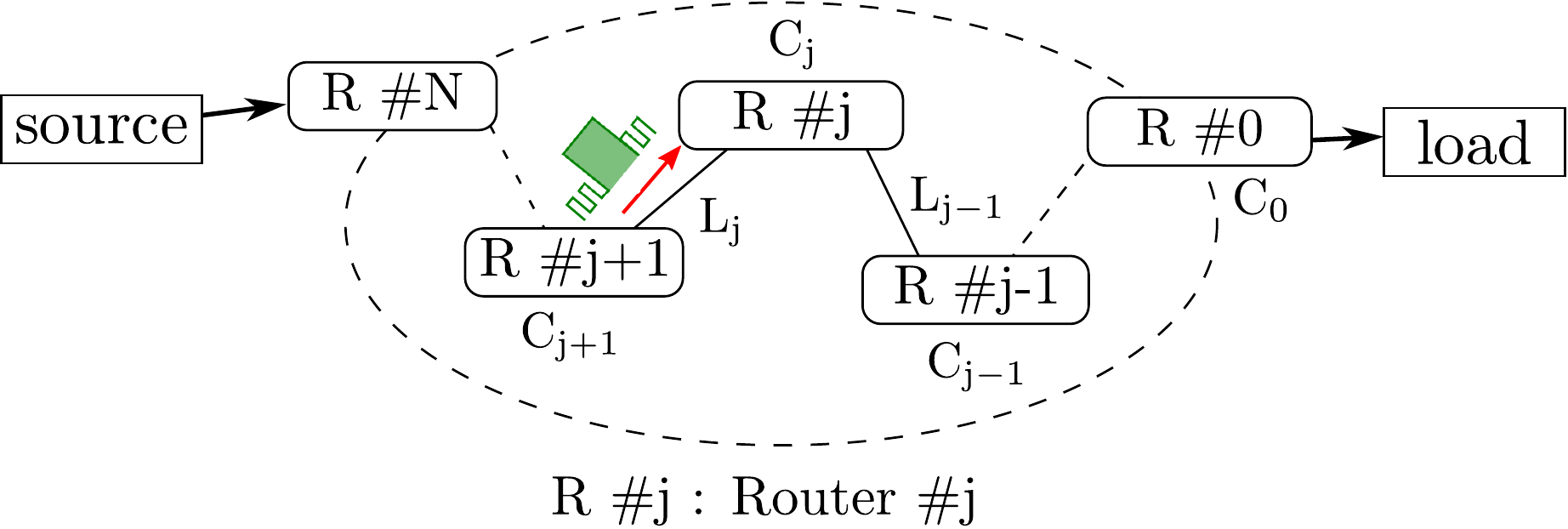}
 \caption{Concept of cascaded power packet dispatching network with N connections.}
 \label{fig6:casNet}
\end{figure}

\begin{figure}
 \centering
 \begin{minipage}{0.45\hsize}
  \centering
  \includegraphics[width = .8\hsize]{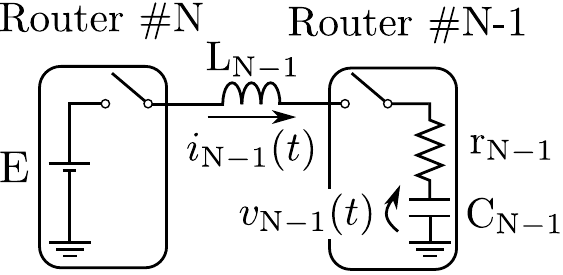}
  \caption{Schematic diagram of a connection \\ between a router and a source.}
  \label{fig:sourceFilt}
 \end{minipage}
 \begin{minipage}{0.45\hsize}
  \centering
  \includegraphics[width = .9\hsize]{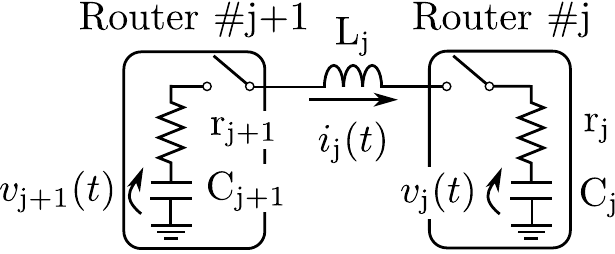}
  \caption{Schematic diagram of a connection \\ between two routers.}
  \label{fig:inNetFilt}
 \end{minipage}
\end{figure}

A concept of the power packet routing between a source and a load is shown in \fref{fig6:casNet}.
Routers connect a source and a load with a cascaded topology.
A connection j ($= 0, 1, \dots$, N-1) consists of routers \# j, \# j+1, and the power line $\rm{L_j}$.
Router \# j has a capacitor ${\rm C_j}$ as storage.
Router \# N is directly connected to the source and does not have storage.
$\rm{L_j}$ also shows the self-inductance of the line.
The current on $\rm{L_j}$ is represented by $i_{\rm j}(t)$.
Firstly, we analyze the power transmission between two routers.
The circuits relevant to power transmission are classified into two circuits shown in \fsref{fig:sourceFilt} and \ref{fig:inNetFilt}.
${\rm r_j}$ shows ${\rm C_j}$'s equivalent series resistance (ESR).
The two circuits are analyzed in the following.

\Fref{fig:sourceFilt} shows the connection with a source.
It has a similar circuit configuration to a buck converter.
When the switches are ON, it can be regarded as an RLC filter.
Thus, during the power transmission, the differential equations of \fref{fig:sourceFilt} are,
\begin{align}
    {\rm C_{\rm N-1}} \frac{{\rm d}v_{\rm N-1}(t)}{{\rm d}t} &= i_{\rm N-1}(t), \\
    (2{\rm r_{on}} + {\rm r_{\rm N-1}})i_{\rm N-1}(t) + {\rm L_{\rm N-1}}\frac{{\rm d}i_{\rm N-1}(t)}{{\rm d}t} &= {\rm E} - v_{\rm N-1}(t).
\end{align}
Where ${\rm E}$ denotes the voltage of the source and ${\rm r_{on}}$ corresponds to the on-resistances of the switch.
The power transmission by the payload is regarded as a step input.
The step response of the RLC filter is roughly divided into overdamped and underdamped.
The response is classified by the parameter $\zeta_{\rm N-1}$.
$\zeta_{\rm N-1}$ is calculated as
\begin{equation}
  \zeta_{\rm N-1} = \frac{2{\rm r_{on}} + {\rm r_{\rm N-1}}}{2}\sqrt{\frac{\rm{C_{\rm N-1}}}{\rm{L_{\rm N-1}}}} \label{eq:zeta1}
\end{equation}
When $\zeta_{\rm N-1} < 1$, the step response becomes underdamped.
$\zeta_{\rm N-1} > 1$ is the condition for the overdamped response.
We set $t = $~0 to be the payload beginning.
Given the initial condition, $i_{\rm N-1}(t)$ becomes;
\begin{align}
   i_{\rm N-1}(t) &= \frac{{\rm E}-v_{\rm N-1}(0)}{{\rm L_{\rm N-1}}\omega_{\rm N-1}\sqrt{1-{\zeta_{\rm N-1}}^2}}\exp(-\zeta_{\rm N-1} \omega_{\rm N-1} t)\sin (\omega_{\rm N-1} \sqrt{1-{\zeta_{\rm N-1}}^2}t) & \textrm{for }\zeta_{\rm N-1} < 1, \label{eq:i1} \\
    i_{\rm N-1}(t) &= \frac{{\rm E}-v_{\rm N-1}(0)}{{\rm L_{\rm N-1}}\omega_{\rm N-1}\sqrt{{\zeta_{\rm N-1}}^2-1}}\exp(-\zeta_{\rm N-1} \omega_{\rm N-1} t)\sinh (\omega_{\rm N-1} \sqrt{{\zeta_{\rm N-1}-1}^2}t) & \textrm{for }\zeta_{\rm N-1} > 1. \label{eq:ih1}
\end{align}
Where $\omega_{\rm N-1}$ is represented by;
\begin{equation}
  \omega_{\rm N-1} = \frac{1}{\sqrt{\rm{L_{\rm N-1}C_{\rm N-1}}}}. \label{eq:omega1}
\end{equation}

\Fref{fig:inNetFilt} shows a connection between two routers in the network.
The differential equations during the payload are;
\begin{align}
    {\rm C_{\rm j}} \frac{{\rm d}v_{\rm j}(t)}{{\rm d}t} &= i_{\rm j}(t), \\
    {\rm C_{\rm j+1}} \frac{{\rm d}v_{\rm j+1}(t)}{{\rm d}t} &= - i_{\rm j}(t), \\
  (2{\rm r_{on}} + {\rm r_{\rm j+1}} + {\rm r_{\rm j}}) i_{\rm j}(t) + {\rm L_j}\frac{{\rm d}i_{\rm j}(t)}{{\rm d}t} &= v_{\rm j+1}(t) - v_{\rm j}(t).
\end{align}
By setting $\omega_{\rm j}$ and $\zeta_{\rm j}$ as follows;
\begin{align}
  \zeta_{\rm j} &= \frac{2{\rm r_{on}}+{\rm r_j}+{\rm r_{j+1}}}{2}\sqrt{\frac{{\rm C_jC_{j+1}}}{{\rm L_j}({\rm C_j}+{\rm C_{j+1}})}} \\
  \omega_{\rm j} &= \sqrt{\frac{({\rm C_j}+{\rm C_{j+1}})}{{\rm L_jC_jC_{j+1}}}}
\end{align}
$i_{\rm j}(t)$ is given as;
\begin{align}
 i_{\rm j}(t) &= \frac{v_{\rm j+1}(0)-v_{\rm j}(0)}{{\rm L_{\rm j}}\omega_{\rm j}\sqrt{1-{\zeta_{\rm j}}^2}}\exp(-\zeta_{\rm j} \omega_{\rm j} t)\sin (\omega_{\rm j} \sqrt{1-{\zeta_{\rm j}}^2}t) &  \textrm{for }\zeta_{\rm j} < 1, \label{eq:i0} \\
 i_{\rm j}(t) &= \frac{v_{\rm j+1}(0)-v_{\rm j}(0)}{{\rm L_{\rm j}}\omega_{\rm j}\sqrt{{\zeta_{\rm j}}^2-1}}\exp(-\zeta_{\rm j} \omega_{\rm j} t)\sinh (\omega_{\rm j} \sqrt{{\zeta_{\rm j}}^2-1}t) &  \textrm{for }\zeta_{\rm j} > 1. \label{eq:ih0}
\end{align}
When the load current is negligible during the payload, \esref{eq:i1}, (\ref{eq:ih1}), (\ref{eq:i0}), and (\ref{eq:ih0}) are also valid if the storage is directly connected to the load.
With these equations, we can analyze the power transmission between routers.

\subsection{Power transmission on multiple connections}\label{sec:twoCon}

\begin{figure}[t]
 \centering
 \includegraphics[width = .7\hsize]{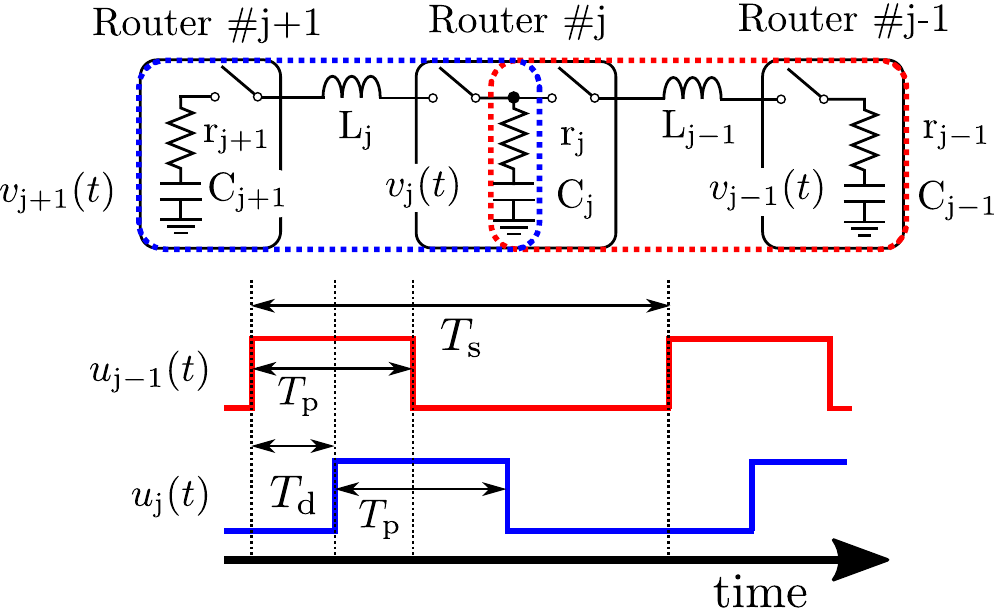}
 \caption{Schematic diagram of cascaded power packet dispatching network with two connections.}
 \label{fig:inTwo}
\end{figure}

Power transmission on two connections is analyzed.
\Fref{fig:inTwo} shows a schematic of the cascaded power packet dispatching network with two connections.
The two connections can transmit power following the switching functions $u_{\rm j-1}(t)$ and $u_{\rm j}(t)$.
They can have different switching cycles and pulse widths.
For simplicity, we assume that $u_{\rm j-1}(t)$ and $u_{\rm j}(t)$ have the same switching cycle $T_{\rm s}$ and pulse width $T_{\rm p}$.
There is another parameter $T_{\rm d}$.
It shows the time difference between the rising edges of $u_{\rm j-1}(t)$ and $u_{\rm j}(t)$.
The effects of $T_{\rm d}$ were experimentally examined in \cite{Katayama2021a}.
The experimental results showed that the storage voltages do not depend on $T_{\rm d}$ when $T_{\rm p} < T_{\rm d} < T_{\rm s}-T_{\rm p}$.
When $T_{\rm d}$ is in the range, the power transmission can be analyzed with the equations in \scref{sec:analysis} since the two connections do not transmit power at the same time.
Then, we consider $T_{\rm d} = 0$ as an example of simultaneous connections.
We assume that $i_{\rm j-1}$ and $i_{\rm j}$ represent the current on ${\rm L_{j-1}}$ and ${\rm L_j}$, respectively.
The current from j+1 to j and j to j-1 is set to plus.
The circuit equations in the case of $T_{\rm d} = 0$ are described as;

\begin{equation}
   \frac{\rm{d}}{{\rm d}t}
    \begin{pmatrix}
      v_{\rm j-1} \\
      v_{\rm j} \\
      v_{\rm j+1} \\
      i_{\rm j-1} \\
      i_{\rm j}
    \end{pmatrix} = \bm{M}
      \begin{pmatrix}
        v_{\rm j-1} \\
        v_{\rm j} \\
        v_{\rm j+1} \\
        i_{\rm j-1} \\
        i_{\rm j}
      \end{pmatrix}. \label{eqa:3mat}
\end{equation}
\begin{equation}
    {\bm M} = \begin{pmatrix}
      0 & 0 & 0 & 1/{\rm C_{j-1}} & 0 \\
      0 & 0 & 0 & -1/{\rm C_j} & 1/{\rm C_j} \\
      0 & 0 & 0 & 0 & -1/{\rm C_{j+1}} \\
      -1/{\rm L_j} & 1/{\rm L_j} & 0 & -(2{\rm r_{on}} + {\rm r_{j-1}} + {\rm r_j})/{\rm L_j} & {\rm r_j}/{\rm L_j} \\
      0 & -1/{\rm L_{j+1}} & 1/{\rm L_{j+1}} & {\rm r_j}/{\rm L_{j+1}} & -(2{\rm r_{on}} + {\rm r_j}+{\rm r_{j+1}})/{\rm L_{j+1}}
    \end{pmatrix}.
\end{equation}

We can find that \eref{eqa:3mat} does not have a unique solution.
Therefore, the energy conservation law is additionally required for analysis.
Furthermore, the unique solution on multiple connections is generally more difficult to find than the one-to-one connection.
Therefore, the power transmission analysis on multiple simultaneous connections requires circuit simulation.

\section{Analytical estimation of storage voltages}\label{sec:estim}

The theoretical examination in \scref{sec:analysis} shows that one-to-one connections allow theoretical power transmission analysis, including the transient response.
Among the power transmission conditions, we focus on the underdamped state.
In the following, we analyze power transmission with the underdamped condition.
Before analyzing the power transmission in the network, we experimentally verify the payload transmission with the underdamped condition.

There are two reasons for the underdamped condition.
The first one is that the current periodically becomes zero.
The second is capacitance.
The zero-current contributes to suppressing the surge voltage at the end of the payload.
When the line current follows \eqref{eq:i1} or (\ref{eq:i0}), it becomes zero at $T_{jn} = n\pi/\omega_{\rm j}\sqrt{1-{\zeta_{\rm j}}^2}$.
Especially, $n=1$ is examined for efficiency.
If we set $T_{jn}$ for the payload duration, the sender and receiver routers can complete the payload transmission without surge voltages.
The surge voltage can reach higher than the source voltage.
It may result in the breakdown of the switches.
It also causes switching loss, which is not considered in \scref{sec:analysis}.
By reducing the loss, the payload with the duration $T_{jn}$ is expected to improve the accuracy of the power transmission analysis.
When $2k-1 < n < 2k$ ($k = 1,2,\dots$), the current flows back to ${\rm C_{j+1}}$ from ${\rm C_j}$.
The backflow current reduces the amount of transmitted energy and increases the transmission loss.
Setting $n=1$ avoids backflow current.
For efficiency, we use $n=1$.
We call the payload with the duration $T_{j1}$ \textit{underdamped payload} in the following.

The second reason is the circuit parameters for the underdamped condition.
When the underdamped response is obtained, $\zeta$ is smaller than 1.
Considering the actual device, multi-layer ceramic (MLCC) and aluminum (Al) electrolytic capacitors are available for the payload under consideration.
MLCC has smaller ESR than Al electrolytic capacitors.
However, the available capacitance of MLCC is smaller than Al electrolytic capacitors.
In this sense, the underdamped condition is suitable for adopting MLCC as storage.

\subsection{Experimental verification}\label{sec:ex}

\begin{figure}[t]
 \centering
 \includegraphics[width = \hsize]{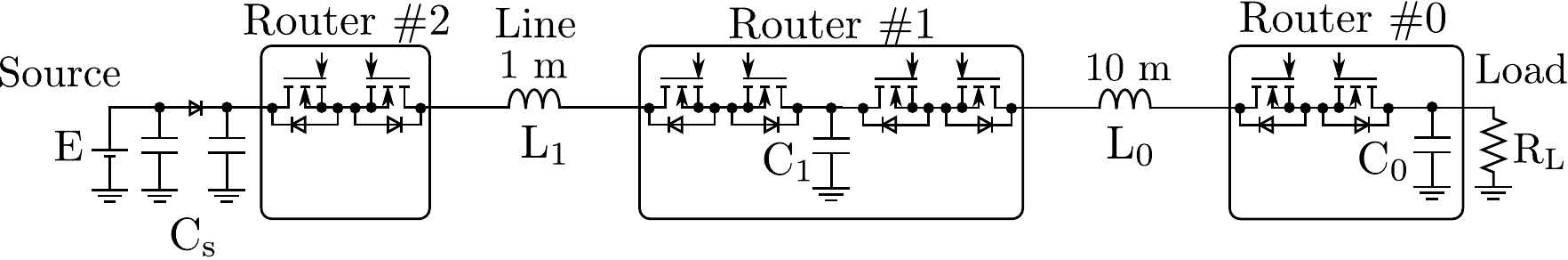}
 \caption{Schematic of an experimental system with three routers, one source, and one load.}
 \label{fig:exThree}
\end{figure}

The underdamped payload transmission is experimentally verified before analyzing the power transmission in the network.
\Fref{fig:exThree} shows the experimental circuit with three routers, one source, and one load.
The power line is a 600-V vinyl insulated vinyl sheathed flat-type (VVF) cable\footnote{Also known as PVC cables.}, popular in Japan for in-home power distribution.
We measured the impedance of the VVF cable of two cores with a core diameter of 1.6~mm by an impedance analyzer.
The inductance of the line is estimated at around 500~nH/m.
The storage device is a multi-layer ceramic capacitor (MLCC) C5750X7S2A106K230KB (TDK).
MLCC has ESR of less than 3.0\,m$\Omega$ at 1.0\,k to 1.0\,MHz frequency range \cite{tdk10uFcap}.
The impedance analyzer also measures the resistances of the line and MLCC.
They are less than 100~m$\Omega$ at the frequency range above.
The capacitance of the storages ${\rm C_0}$ and ${\rm C_1}$ are set at 20~$\mu$F.
It is selected to satisfy the underdamped condition.
Two capacitors ${\rm C_s}$ of 40~$\mu$F are to stabilize the input voltage.
The diode between ${\rm C_s}$ blocks unintentional current to the source.
The detailed configuration and operation of the router are described in \cite{Katayama2020a}.
The signal units are included in the experimental system but omitted in \fref{fig:exThree} for simplicity.
$\zeta_0$ is estimated to be 0.24.
Thus, the underdamped current waveform will be shown on $i_0$.
The estimated pulse width $T_{\rm p0}$ is 22~$\mu$s.
It is calculated from the storage capacitance of 20~$\mu$F, estimated inductance 5.0~$\mu$H, typical on-resistances of the metal-oxide-semiconductor field-effect transistor (MOSFET) \cite{Co2018}, and the measured resistances of 100~m$\Omega$.
Whether $i_1$ satisfies the underdamped condition is not easily identified since a diode is implemented.
The payload width $T_{\rm p0}$ and $T_{\rm p1}$ were changed manually to achieve zero-current turn-off.
The time step was 1.0~$\mu$s due to the hardware rating \cite{Katayama2020a}.
Both connections in the experimental system had the same switching period.
The period can be set at any duration if a router gets connected to another router at a time.
We set the period so that the duty cycle becomes 18\% on the connection between Router\,\#0 and \#1.
We measured the power line current $i_0$ and $i_1$, and the storage voltage $v_{0}$ and $v_{1}$.

\begin{figure}[t]
  \centering
  \includegraphics[width = .8\hsize]{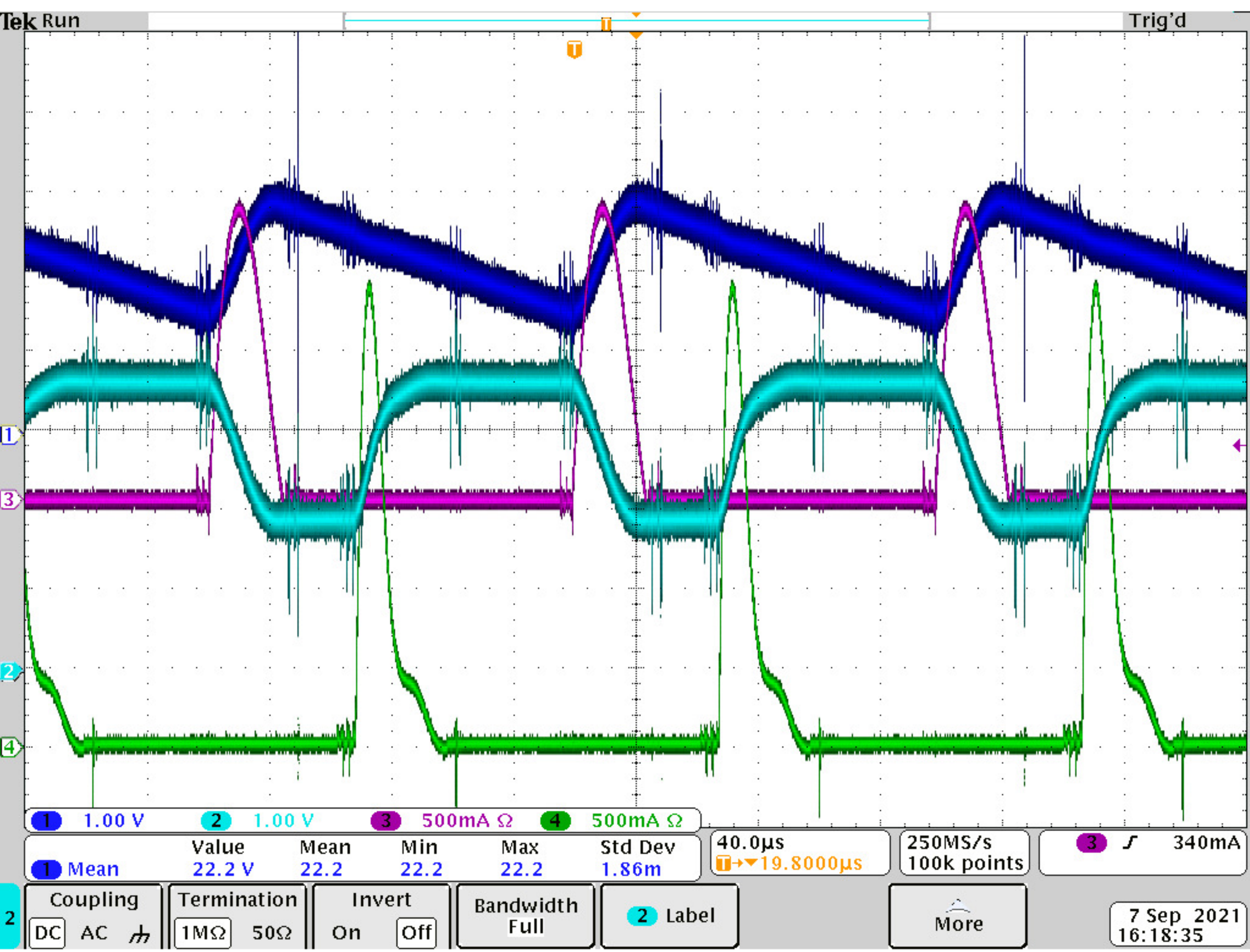}
  \caption{Experimental results of $v_{0}$ (CH1), $v_{1}$ (CH2), $i_0$ (CH3), and $i_1$ (CH4). $v_{0}$ and $v_{1}$ are measured with an offset voltage of 20.0V.}
  \label{fig:exCurVol}
\end{figure}

The measured experimental waveforms are shown in \fref{fig:exCurVol}.
$i_0$ shows the underdamped current waveform.
Its pulse duration $T_{\rm p0}$ of 26~$\mu$s is longer than the estimated duration of 22~$\mu$s.
The difference owes to the assumption that the MOSFETs change their states instantly, and the load is negligible.
The characteristics of MLCC against DC voltage bias is another reason \cite{tdk10uFcap}.
$i_1$ also becomes zero when $T_{\rm p1}$ is set at 31~$\mu$s.
It is affected by the diode and did not take a similar waveform to $i_0$.
$v_0$ and $v_1$ include short pulses.
They are due to the signal part of the power packet.

$v_1$ shows that the storage in the network takes two voltages when isolated from other storage.
We denote these voltages with $v_{\rm jH}$ and $v_{\rm jL}$ for the storage ${\rm C_j}$.
They correspond to the highest and lowest storage voltages at a periodic steady state.
By transmitting (or receiving) an underdamped payload, the storage voltage shifts from $v_{\rm jH}$ ($v_{\rm jL}$) to $v_{\rm jL}$ ($v_{\rm jH}$).
The shift is analyzed with \esref{eq:i1} and (\ref{eq:i0}).
The initial conditions of these equations are the storage voltages before transmission.
Therefore, $v_{\rm jH}$ and $v_{\rm jL}$ are necessary to analyze the power transmission in the network.

\subsection{Relationships between storage voltages}\label{sec:anaNet}

The experimental results imply that the storage voltages can be analyzed with their highest and lowest voltages $v_{\rm jH}$ and $v_{\rm jL}$ at periodic steady states.
We consider the underdamped payload transmission in the circuit shown in \fsref{fig:sourceFilt} and \ref{fig:inNetFilt}.
We can calculate storage voltages after an underdamped payload transmission with \esref{eq:i1} and (\ref{eq:i0}) .
The storage voltages before and after transmission correspond to their highest or lowest voltages $v_{\rm jH}$ and $v_{\rm jL}$.
In \fref{fig:sourceFilt}, ${\rm C_{N-1}}$ receives an underdamped payload.
By receiving the payload, its storage voltage shifts from $v_{\rm{N-1L}}$ to $v_{\rm{N-1H}}$.
Considering \eref{eq:i1}, the relationship between $v_{\rm{N-1L}}$ and $v_{\rm{N-1H}}$ is shown as;
\begin{equation}
  v_{\rm{N-1H}} = v_{\rm{N-1L}} + ({\rm E}-v_{N-1 \rm{L}})\frac{e_{\rm N-1}}{{\rm C_{N-1}}} \label{eq:v1te}.
\end{equation}
For simplicity, we put $e_{\rm N-1}$ as;
\begin{equation}
  e_{\rm N-1} = \frac{1}{\rm{L_{N-1}}{\omega_{\rm N-1}}^2}\left\{1 + \exp \left( -\frac{\zeta_{\rm N-1}\pi}{\sqrt{1-{\zeta_{\rm N-1}}^2}} \right) \right\}.
\end{equation}
Likewise, an underdamped payload transmission in \fref{fig:inNetFilt} shifts $v_{\rm{j+1H}}$ to $v_{\rm{j+1L}}$ and $v_{\rm jL}$ to $v_{\rm jH}$.
They have the expressions;
\begin{align}
  v_{\rm{j+1L}} = v_{\rm{j+1H}} - (v_{\rm{j+1H}}-v_{\rm jL})\frac{e_{\rm j}}{{\rm C_{j+1}}}, \label{eq:v1t} \\
  v_{\rm jH} = v_{\rm jL} + (v_{\rm{j+1H}}-v_{\rm jL})\frac{e_{\rm j}}{{\rm C_j}}. \label{eq:v0t}
\end{align}
\Esref{eq:v1te} to (\ref{eq:v0t}) show that the storage voltages before and after transmission have linear relationships.
Therefore, by combining them, the storage voltages at periodic steady states are estimated theoretically.

Let us consider a cascaded network with N connections, as shown in \fref{fig6:casNet}.
The connection \#i consists of routers \#i and \#j+1, and the power line \#i.
For the cascaded network with ${\rm N}$ connections, the combination of \esref{eq:v1te} to (\ref{eq:v0t}) is arranged as;
\begin{equation}
  \bm{M_{\rm N}}{\bm v} = \bm{e_{\rm{s}}}. \label{eq:distEq}
\end{equation}
Where, $\bm{v}$ and $\bm{e_{\rm{s}}}$ are vectors of $2{\rm N}$ elements.
\begin{align}
  \bm{v} &= (v_{0\rm{L}}, v_{0\rm{H}}, v_{1\rm{L}}, \dots, v_{\rm N-1L}, v_{\rm N-1H})^T, \\
  \bm{e_{\rm{s}}} &= (v_{0\rm{L}}, 0, 0, \dots, 0,  e_{\rm{N-1}}{\rm E}/{\rm C_{N-1}})^T.
\end{align}
$\bm{M_{\rm N}}$ is a $2{\rm N} \times 2{\rm N}$ matrix.
The components of $\bm{M_{\rm N}}$ are written with index $i = 0, 1, \dots, {\rm N-1}$ as follows;
\begin{align}
  M_{\rm 2j+2,2j+1} &= \frac{1}{{\rm L_jC_{i}}{\omega_{\rm j}}^2}\left\{1 + \exp \left( -\frac{\zeta_{\rm j}\pi}{\sqrt{1-{\zeta_{\rm j}}^2}} \right) \right\}-1, \\
  M_{\rm 2j+3,2j+1} &= -\frac{1}{{\rm L_jC_{j+1}}{\omega_{\rm j}}^2}\left\{1 + \exp \left( -\frac{\zeta_{\rm j}\pi}{\sqrt{1-{\zeta_{\rm j}}^2}} \right) \right\}, \\
  M_{\rm 2j+2,2j+4} &= -\frac{1}{{\rm L_jC_{i}}{\omega_{\rm j}}^2}\left\{1 + \exp \left( -\frac{\zeta_{\rm j}\pi}{\sqrt{1-{\zeta_{\rm j}}^2}} \right) \right\}, \\
  M_{\rm 2j+3,2j+4} &= \frac{1}{{\rm L_jC_{j+1}}{\omega_{\rm j}}^2}\left\{1 + \exp \left( -\frac{\zeta_{\rm j}\pi}{\sqrt{1-{\zeta_{\rm j}}^2}} \right) \right\}-1, \\
  M_{\rm 2j+1,2j+1}, M_{\rm 2j+2,2j+2} &= 1.
\end{align}
The other components of $\bm{M_{\rm N}}$ are 0.
For example, $\bm{M_3}$ becomes;
\begin{align}
  \bm{M_3} =
  \begin{pmatrix}
    1 & 0 & 0 & 0 & 0 & 0 \\
    -1+e_0/\rm{C_0} & 1 & 0 & -e_0/\rm{C_0} & 0 & 0 \\
    -e_0/\rm{C_1} & 0 & 1 & -1+e_0/\rm{C_1} & 0 & 0 \\
    0 & 0 & -1+e_1/\rm{C_1} & 1 & 0 & -e_1/\rm{C_1}  \\
    0 & 0 & -e_1/\rm{C_2} & 0 & 1 & -1+e_1/\rm{C_2} \\
    0 & 0 & 0  & 0 & -1+e_2/\rm{C_2} & 1 \\
  \end{pmatrix}
  .
\end{align}
$\bm{M_{\rm N}}$ is calculated from the circuit components.
It does not depend on the circuit state.
When $\bm{M_{\rm N}}$ is non-singular, \eref{eq:distEq} has a unique solution.
The solution estimates the storage voltages at a periodic steady state.
The power transmission analysis in the network is executable with \esref{eq:i1} and (\ref{eq:i0}), and the estimated storage voltages.

\subsection{Numerical verification of voltage estimation}\label{sec:verify}

\begin{figure}[!t]
 \centering
 \includegraphics[width = .8\hsize]{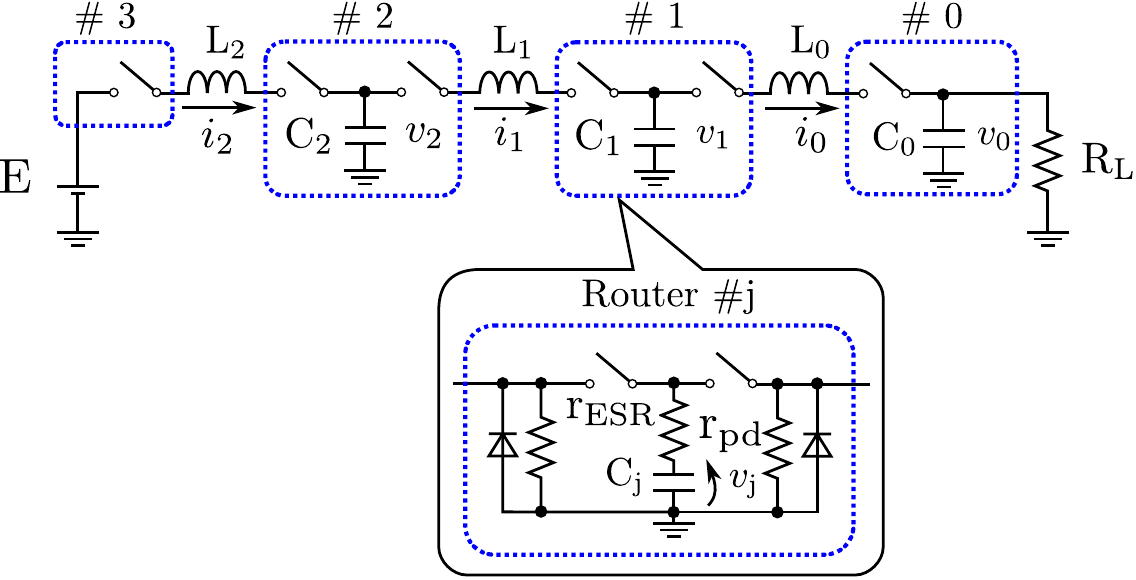}
 \caption{Schematic of cascaded network with three connections. The detailed diagram of the router is shown below the network.}
 \label{fig:schSim}
\end{figure}

\begin{table}[!t]
 \caption{Simulation setup.}
 \label{tab:const}
 \begin{center}
  \begin{tabular}{cccc}\hline
    Name & Symbol & Model & Value \\ \hline
    Source voltage & E & DC source & 24.0~V \\
    Load resistance & ${\rm R_L}$ & Resistor & 100~$\Omega$ \\
    Storage &  ${\rm C_0}$, ${\rm C_1}, {\rm C_2}$ & Capacitor & 1.0,~10,~20,~40~$\mu$F \\
    Capacitor ESR & ${\rm r_{ESR}}$ & -- & 100~m$\Omega$ \\
    Line & ${\rm L_0}, {\rm L_1}, {\rm L_2}$ & Inductor & 1.0~$\mu$H \\
    MOSFETs &  & SCT3022AL & -- \\
    Freewheeling Diode & & UF4002 & -- \\
    Pulldown Resistance & ${\rm r_{pd}}$ & Resistor & 30.0~k$\Omega$ \\
    Switching Cycle & $T_{\rm s}$ & - &  \\ \hline
    \end{tabular}
 \end{center}
\end{table}

The estimation accuracy by \eref{eq:distEq} is examined with the circuit simulation.
\Fref{fig:schSim} shows the schematic of the cascaded router network with three connections.
The circuit model is developed in LTSpice XVII.
The device models and simulation parameters are shown in \tref{tab:const}.
A freewheeling diode and resistor are placed at each port of routers instead of the signal unit.
The resistor pulls down the power line voltage to the ground level during the interval of power transmission.
They simplify the simulation model and shorten the simulation time.
It is confirmed beforehand that the replacement has little impact on the results, such as $T_{\rm pi}$ and transmitted power.
$v_{\rm 0L}$ is set at 21.6~V, 90~\% of E.
The duration of the underdamped payload is calculated for each connection.
In the simulation, the duration of $T_{\rm j1}$ is represented by the bit length of the payload.
The duration of one-bit ${\rm t_u}$ is set at 1.0~$\mu$s \cite{Katayama2020a}.
The bit length $B_{\rm j1}$ of $T_{\rm j1}$ is set to satisfy the following condition;
\begin{equation}
  B_{\rm j1}{\rm t_u} \le T_{\rm j1} < (B_{\rm j1} + 1){\rm t_u}.
\end{equation}
The switching cycle $T_{\rm s}$ is calculated from the estimated $T_{01}$, $v_{\rm 0L}$, $v_{\rm 0H}$, ${\rm R_L}$, ${\rm C_0}$, and ESR of ${\rm C_0}$.
$T_{\rm s}$ satisfies the equation;
\begin{equation}
  T_{\rm s} = T_{01} + {\rm C_0}({\rm R_L} + {\rm r_{ESR}})\log \frac{v_{\rm 0H}}{v_{\rm 0L}}.
\end{equation}
In the simulation, $T_{\rm s}$ is also realized by bit length $B_{\rm s}$ like $T_{j1}$.
The simulation ran for 10~ms and the averaged storage voltages were calculated with the last two periods.
The 10~ms is long enough for the simulation to reach a periodically stable state.
The initial storage voltages are set at the solution of \eref{eq:distEq}.
The simulation was run several times with different capacities of ${\rm C_0}$, ${\rm C_1}$, and ${\rm C_2}$.

\begin{table}[!t]
 \caption{Comparison of voltages with 20~$\mu$F storage.}
 \label{tab:re20}
 \begin{center}
  \begin{tabular}{c|cccccc}
      & $v_{\rm 0L}$ & $v_{\rm 0H}$ & $v_{\rm 1L}$ & $v_{\rm 1H}$ & $v_{\rm 2L}$ & $v_{\rm 2H}$ \\ \hline
    Estimation & 21.60~V & 22.72~V & 22.34~V & 23.47~V & 23.09~V & 24.21~V \\
    Simulation & 21.40~V & 22.58~V & 22.10~V & 23.49~V & 22.93~V & 24.26~V
    \end{tabular}
 \end{center}
\end{table}

\begin{figure}[!t]
 \centering
 \includegraphics[width = .9\hsize]{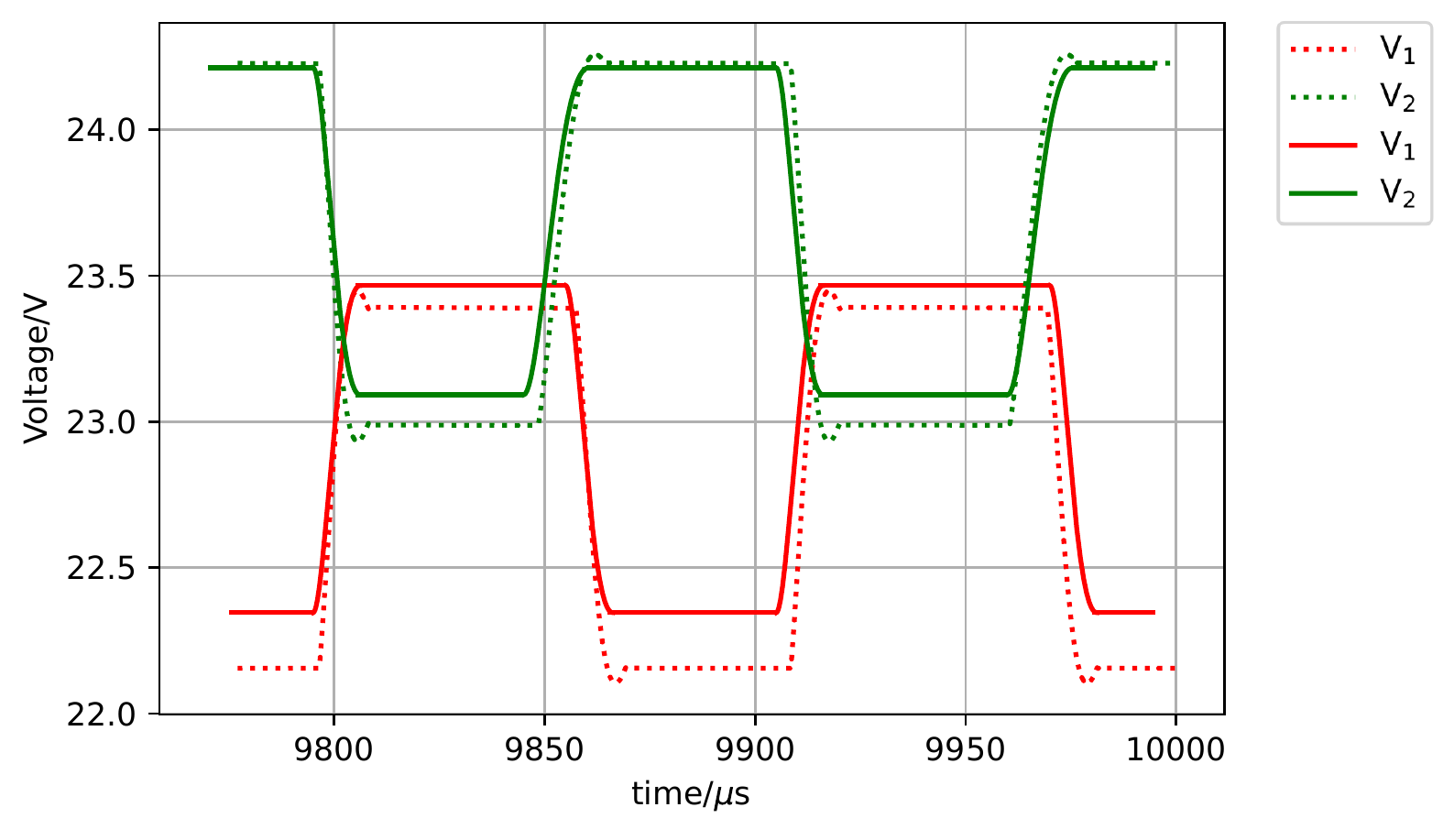}
 \caption{Comparison of voltage waveforms of $v_1(t)$ and $v_2(t)$ with 20~$\mu$F capacitors. Dotted lines are simulation results. Solid lines are reconstructed from estimated storage voltages and \esref{eq:i1} and (\ref{eq:i0}).}
 \label{fig:reSimAlge}
\end{figure}

\begin{table}[!t]
 \caption{Comparison of solution of the modeling and simulation.}
 \label{tab:result}
 \begin{center}
  \begin{tabular}{ccccc}
    ${\rm C_0}$, ${\rm C_1}$, and ${\rm C_2}$ & Averaged $V_{\rm dif}$ & RMSE & $T_{\rm s}$ & $T_{01}$\\ \hline
    1.0~$\mu$F & 2.91~V & 1.93~$\times 10^{-2}$ & 13.98~$\mu$s & 2.23~$\mu$s \\
    10~$\mu$F & 1.58~V & 7.05~$\times 10^{-3}$ & 70.65~$\mu$s & 7.42~$\mu$s \\
    20~$\mu$F & 1.28~V & 6.44~$\times 10^{-3}$ & 112.37~$\mu$s & 11.16~$\mu$s \\
    40~$\mu$F & 1.06~V & 6.28~$\times 10^{-3}$ & 183.10~$\mu$s & 18.36~$\mu$s \\\hline
    \end{tabular}
 \end{center}
\end{table}

\Tref{tab:re20} shows $v_{\rm jH}$ and $v_{\rm jL}$ in the case of 20~$\mu$F storage.
The upper row shows the estimated voltages as the solutions of \eref{eq:distEq}.
\Fref{fig:reSimAlge} shows storage voltages $v_1(t)$ and $v_2(t)$.
The dotted lines are the simulation results, and the solid lines are reconstructed from the estimated voltages.
Though the theoretical analysis does not consider the dynamics of the switches, they are in a good match.
The voltage difference between $v_{\rm jH}$ and $v_{\rm jL}$ is almost the same in the three routers.
It is because the same storage capacitances are used.
In addition, $v_{2{\rm H}}$ is higher than E.
It is also one of the characteristics of an underdamped response.
$v_{2{\rm H}}$ increases as the voltage difference between $v_{2{\rm H}}$ and $v_{2{\rm L}}$ increases.

The results of other parameters are summarized in \tref{tab:result}.
For each case, the averaged voltage difference $V_{\rm dif}$ between $v_{\rm jH}$ and $v_{\rm jL}$ is calculated from the simulation results.
The results show that $V_{\rm dif}$ decreases as the storage capacitance increases.
It implies that the storage connected to the load should have a large capacitance to reduce the voltage ripple.
The root mean square errors (RMSEs) between the estimated voltages and simulation results are calculated.
Each $v_{\rm jH}$ and $v_{\rm jL}$ are normalized by ${\rm E}$.
Then, RMSE is calculated from the normalized difference between the estimation and simulation.
The RMSE decreases as the capacitance get larger.
In the case of 1.0~$\mu$F, RMSE is more than two times as large as the 10~$\mu$F case.
The time difference between $B_{\rm j0}{\rm t_u}$ and $T_{\rm j0}$ against $T_{\rm j0}$ increases as the storage capacitance decreases.
It indicates that the voltage estimation accuracy depends on time accuracy.

The advantage of \eref{eq:distEq} is the required time to estimate voltages.
For example, \eref{eq:distEq} with three connections can be solved in a second by a computer.
The 10~ms-simulation took some minutes by the same machine.
The network with three connections is the second smallest one.
The time to reach steady-state depends on the simulation environment and the number of routers.
The estimation method is useful to analyze power transmission in the network with multiple connections.

\section{Conclusion}\label{sec:conclusion}

This paper proposes an analytical voltage estimation to design power transmission with power packets.
Firstly, the power transmission between two routers is analyzed as an RLC filter.
Secondly, the power transmission on two connections is investigated.
These analyses reveal that the power transmission between two routers allows theoretical analyses.
We focused on an underdamped condition for surge suppression and capacitance of the storage.
The underdamped payload transmission is verified by the experiment.
The experimental results show that the storage voltages at a periodic steady state can be analyzed with their highest and lowest voltages.
Furthermore, the linear relationships between storage voltages before and after sending/receiving the payload are found with the underdamped payload transmission.

In addition, an analytical storage voltage estimation method is proposed.
It sets each storage's highest and lowest voltages at a periodic steady state.
Linear equations give the relationships between these voltages.
They are solved with the minimum load and source voltages.
Estimated storage voltages are compared with the circuit simulation by LTSpice XVII.
Both estimations are in a good match when the storage capacity is larger than 10~$\mu$F.
In the case of 1.0~$\mu$F, the duration of the underdamped payload realized by the bit length is not accurate enough to estimate storage voltages as good as the cases with larger capacitances.
Further studies on the capacity are necessary to guarantee the accuracy of the estimation.
The estimated voltages also provide the current waveforms and power transmission efficiency.
The power transmission analysis in the network will be available with the proposed method.



\end{document}